\documentclass[10pt,letterpaper]{article}
\usepackage{opex3}
\usepackage[latin1]{inputenc}
\usepackage{amssymb,amsmath}
\usepackage{cite}

\newcommand{\rmi}{{\rm{i}}}
\newcommand{\rme}{{\rm{e}}}
\newcommand{\rmd}{{\rm{d}}}

\begin{document}

\title{Diffraction effects in length measurements by laser interferometry}

\author{C.\ P.\ Sasso, E.\ Massa and G.\ Mana}
\address{INRIM -- Istituto Nazionale di Ricerca Metrologica, str.\ delle Cacce 91, 10135 Torino, Italy}
\email{$^*$e.massa@inrim.it} 


\begin{abstract}
High-accuracy dimensional measurements by laser interferometers require corrections because of diffraction, which makes the effective fringe-period different from the wavelength of a plane (or spherical) wave $\lambda_0$. By using a combined X-ray and optical interferometer as a tool to investigate diffraction across a laser beam, we observed wavelength variations as large as $10^{-8}\lambda_0$. We show that they originate from the wavefront evolution under paraxial propagation in the presence of wavefront- and intensity-profile perturbations.
\end{abstract}


\bibliographystyle{osajnl}
\bibliography{wavefront}

\section{Introduction}
In high-accuracy dimensional measurements by laser interferometry, one of the largest corrections required is due to diffraction. Examples are the measurements of Si lattice-parameter by combined X-ray and optical interferometry \cite{Siegert:1981,Bergamin:1994,Bergamin:1997,Massa:2015}, of gravity by free-fall gravimeters \cite{Westrum:2003,Lennart:2007,D'Agostino:2011}, and of diameters by Fizeau interferometers \cite{Nicolaus:2009,Bartl:2011,Kuramoto:2011,Birk:2011,Birk:2012,Andreas:2015}. At the $10^{-9}$ level of relative accuracy it is not possible to trace back wavelength to frequency via the plane-wave dispersion equation. In fact, the laser beam spreads outside the region in which it would be expected to remain in plane wave propagation, the wave fronts bend, and their spacing varies from one point to another and it is different from the wavelength of a plane wave. As a result, measurements must be corrected.

If the interfering beams are paraxial and coaxial, provided that the wavefront phase-shift is small with respect to the Rayleigh distance, extensive investigations led to analytical corrections \cite{Monchalin:1981,Dorenwendt:1976,Mana:1989,Bergamin:1999,Westrum:2003,Lennart:2007}. If the beams are Gaussian, analytical corrections are known also for small (with respect to the beam divergence and radius) wavefront misalignments and shears \cite{Cavagnero:2006}.

In order to assess the accuracy of the interference model, a $5\times 5$ photodiode matrix was used to carry out separate measurements of the same displacement of an X-ray interferometer in 25 points of the beam wavefront. The measured displacements pictured unexpected differences. In this paper, we report about the apparatus, the measurements, and the anomalies observed. Next, we describe a model based on the paraxial propagation of irregular beam-profile and -wavefront, which explains the observations made.

\section{Combined X-ray and optical interferometry}
\begin{figure}
\centering
\includegraphics[height=70mm]{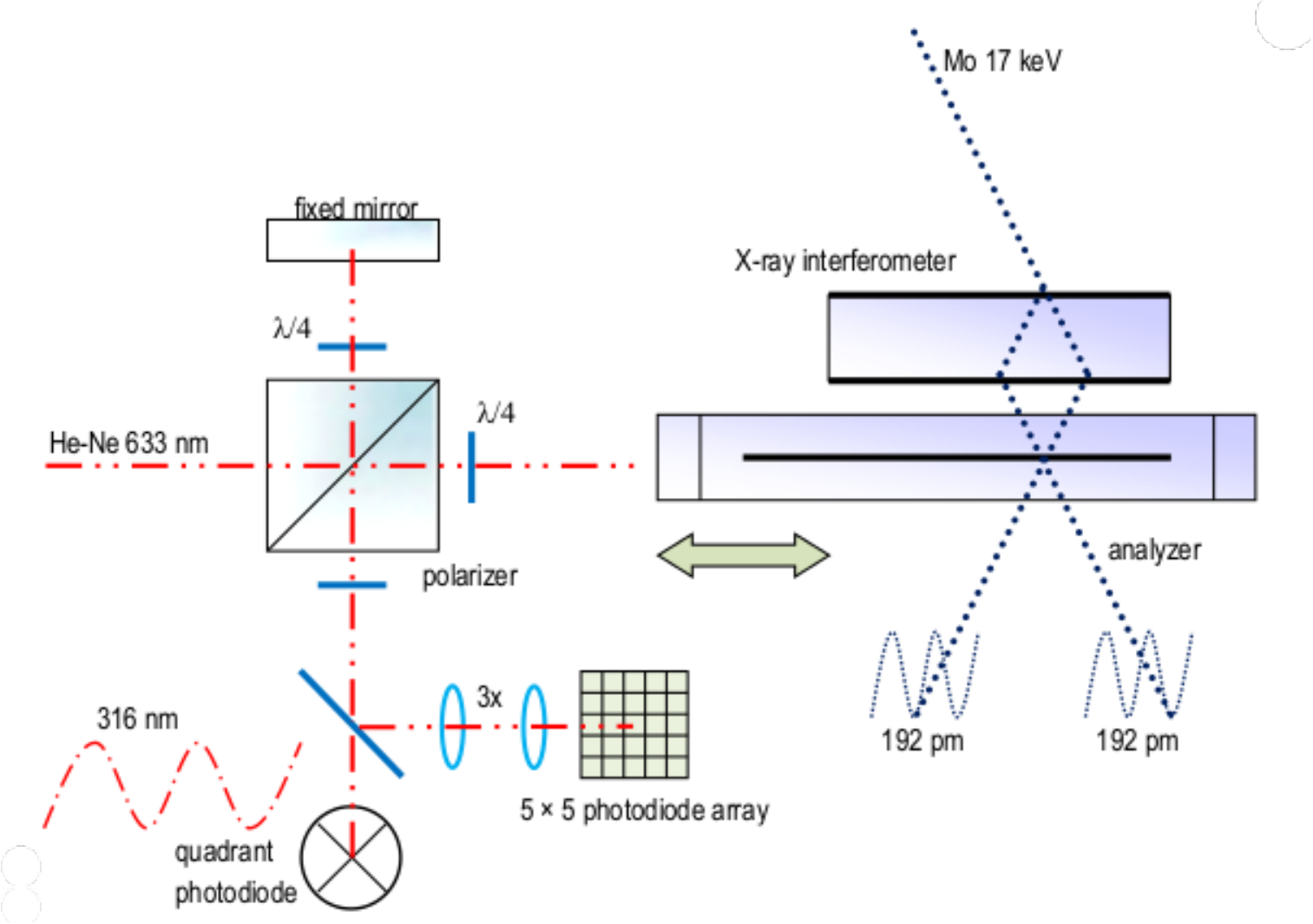}
\caption{Schematics of the combined X-ray and optical interferometer. The analyzer displacement is simultaneously measured in terms of X-ray and optical fringes; the quadrant detector monitor the alignment of the interfering wavefronts.}\label{xroi}
\end{figure}

As shown in Fig.\ \ref{xroi}, a combined X-ray and optical interferometer consists of three Si crystals cut so that the (220) diffracting  planes are orthogonal to the crystal surfaces. X rays from a 17 keV Mo K$\alpha$ source having a $(10\times 0.1)$ mm$^2$ line focus are split by the first two crystals and recombined by the third, which is called the analyzer. The interference pattern is imaged onto a multianode photomultiplier tube through a pile of eight NaI(Tl) scintillator crystals.

When the analyzer is moved along a direction orthogonal to the (220) planes, a periodic variation of the transmitted and diffracted X-ray intensities is observed, the period being the diffracting-plane spacing, $d$. The analyzer embeds a front mirror, so that its displacement is measured by optical interferometry, where the output signal is integrated over the whole interference pattern. The necessary picometer resolution is achieved by polarization encoding and phase modulation. To eliminate the adverse influence of the refractive index of air and temperature, the measurement is carried out in a vacuum and the temperature is controlled up to millikelvin stability and uniformity. In order to ensure the traceability to the meter, the frequency of the laser source is locked to a transition of the $^{127}$I$_2$ molecule.

The interferometer operates over displacements up to 5 cm. This capability is obtained by means of a carriage sliding on a quasi-optical rail. An active tripod with three piezoelectric legs rests on the carriage. Each leg expands vertically and shears in the transverse directions, thus allowing compensation for the sliding errors and positioning to atomic-scale accuracy. Parasitic rotations and transverse motions are sensed via laser interferometry and capacitive transducer; feedback loops provide picometer positioning, nanoradian alignment, and movement with nanometer straightness \cite{Bergamin:1993,Bergamin:2003}.

The measurement equation is $d = m\lambda/(2n)$ where $n$ is the number of X-ray fringes in $m$ optical fringes of $\lambda/2$ period. In practice, $d$ is determined by comparing the periods of the X-ray and optical fringes. Starting from an approximate value, this is done by measuring the X-ray fringe fraction at the ends of increasing displacements and updating the $\lambda/(2d)$ value at each step. The measurement resolution approaches $10^{-9}d$, which means that the experiment is sensitive to $10^{-9}\lambda$ variations of the beam wavelength.

The period of the interference fringes is not equal to the plane-wave wavelength $\lambda_0$; therefore, measurements are corrected for the difference between the effective period of the integrated signal and $\lambda_0$. Provided the interfering beams are coaxial and the optical-path is much smaller than the Rayleigh length, the effective wavelength differs from the plane wave value by $\rm{Tr}(\Gamma)/2$, where $\Gamma$ is the central second-moment matrix of the beam \cite{Bergamin:1999}. This correction holds for any paraxial beam, no matter it is Gaussian or not. If the beam owns cylindrical symmetry, $\rm{Tr}(\Gamma)/2=\theta_0^2/4$, where $\theta_0$ is the far-field divergence. With a typical 0.2 mrad divergence, this correction amounts to $10^{-8}\lambda_0$. In the case of Gaussian beams, the extension to non coaxial beams can be found in \cite{Cavagnero:2006,Fujimoto:2007}.

\section{Wavelength profile of the laser beam}
\subsection{Experimentation}
\begin{figure}
\centering
\includegraphics[width=75mm]{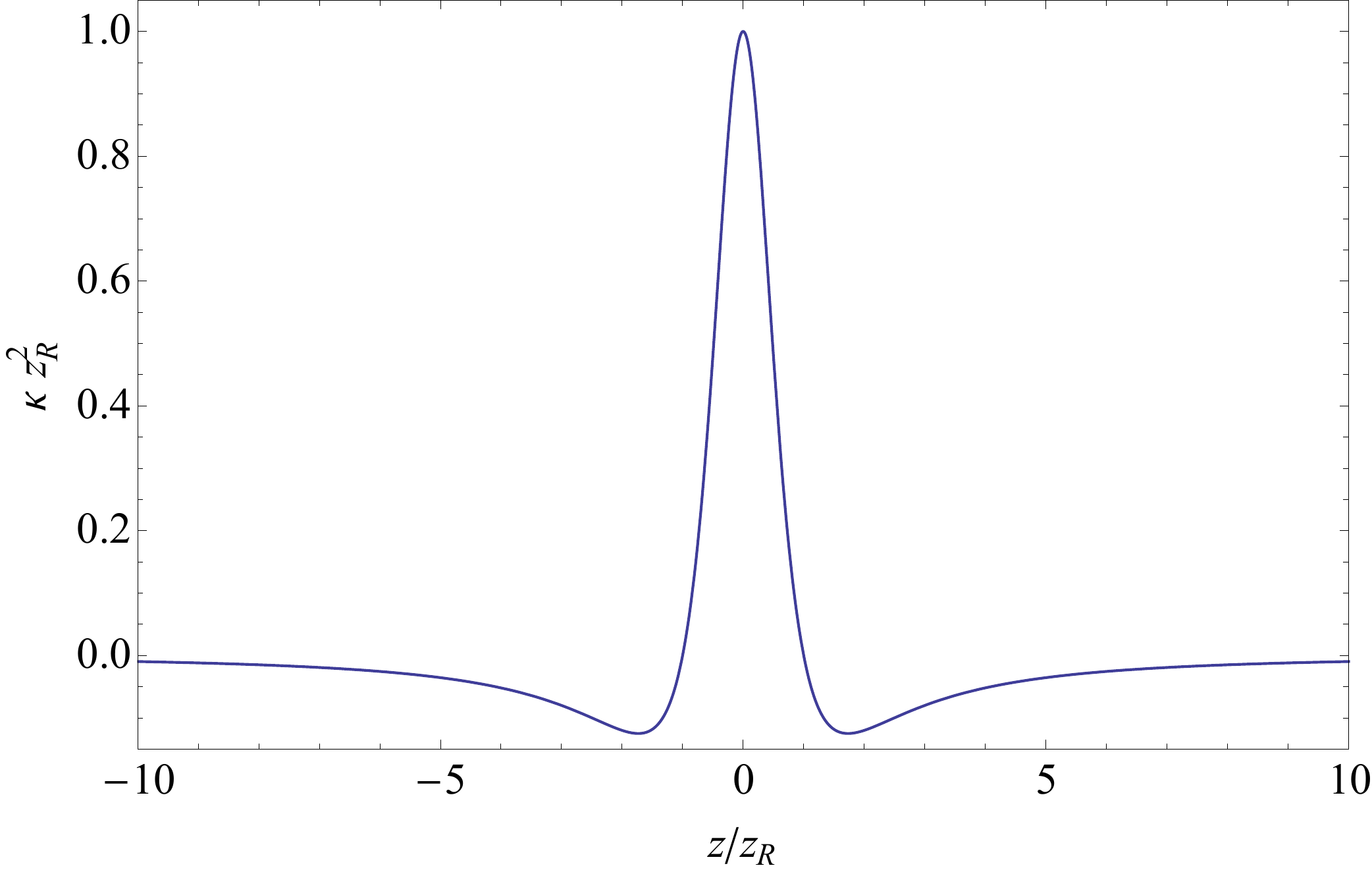}
\caption{Curvature of the wavelength profile across a Gaussian beam.}\label{lambda-profile}
\end{figure}

The phase profile of a Gaussian beam is
\begin{equation}\label{phase:1}
 \varphi(x,y;z) = \frac{k_0(x^2+y^2)}{2R(z)} ,
\end{equation}
where we used the $-\rmi \omega t$ convention for the beam time-evolution, $R(z)=z\big[1 + (z_R/z)^2\big]$ is the wavefront radius-of-curvature, $z_R$ is the Rayleigh distance, $z$ is the distance from the beam waist, $k_0=2\pi/\lambda_0$ is the wave number. $x$ and $y$ are transverse coordinates, and, for the sake of simplicity, we assumed cylindrical symmetry. Accordingly, the wavelength across the beam,
\begin{equation}\label{dl}
 \frac{\Delta \lambda}{\lambda_0} = -\frac{\partial_z \varphi}{k_0} = \frac{(z^2 - z_R^2)(x^2+y^2)}{2(z^2 + z_R^2)^2} ,
\end{equation}
is parabolic with curvature $\kappa = (z^2 - z_R^2)/(z^2 + z_R^2)^2$, which curvature is shown in Fig.\ \ref{lambda-profile}.

From the analysis of the beam parameters given in table \ref{table:1}, the expected sagitta of a 2 mm wavelength profile across the beam is in the $10^{-8}\lambda$ range; therefore, we concluded that the profile curvature should be detectable. Consequently, in order to assess the accuracy of the Gaussian interference models, a square $5\times 5$ photodiode matrix (HAMAMATSU mod. S7585) was used to observe the same interferometer displacement (about 1 mm) in 25 different points across the beam. A $3\times$ beam expander was used to match the interference pattern to the photodiode area, $(7.5 \times 7.5)$ mm$^2$. As shown in Fig.\ \ref{xroi}, to avoid misalignments during the measurements, the analyzer rotation was measured by an optical lever, that is, by detecting the differences between the displacements observed in four points of the interference pattern. A quadrant detectors was used to split the signal integration and the analyzer attitude was electronically controlled to nullify the observed differences and to keep parallel the interfering wavefronts \cite{Bergamin:1993,Bergamin:1994}.

\begin{table}
  \begin{center}
    \caption{Beam parameters and curvature of the best-fit approximations of the wavelength profiles in Fig.\ \ref{hama}. $F$ is the focal length of the fiber collimator, $w_D$ is the beam radius at the detection plane, $z_D$ is the detector distance from the beam waist, $\kappa$ is the observed curvature.} \label{table:1}
    \begin{tabular}{crrrrccrcr}
      \hline
$\#$ &\multicolumn{1}{c}{$F$}             &\multicolumn{1}{c}{$\lambda_0$} &\multicolumn{1}{c}{$\theta_0$}
     &\multicolumn{1}{c}{$z_R$}           &\multicolumn{1}{c}{$w_D$}       &\multicolumn{1}{c}{$z_D\,^a$}
     &\multicolumn{1}{c}{$\kappa$}        &\multicolumn{1}{c}{$z_D\,^b$}   &\multicolumn{1}{c}{$z_D\,^c$} \\
     &\multicolumn{1}{c}{mm}              &\multicolumn{1}{c}{nm}          &\multicolumn{1}{c}{mrad}
     &\multicolumn{1}{c}{m}               &\multicolumn{1}{c}{mm}          &\multicolumn{1}{c}{m}
     &\multicolumn{1}{c}{mm$^{-2}$}       &\multicolumn{1}{c}{m}           &\multicolumn{1}{c}{m}                \\
      \hline
      1 &12 & 632 &0.20 &  5.0 &--  &--          &$-4\times 10^{-9}$    &$\pm 7.0$ & $-2 < z_D < 4$\\
      2 &8  & 532 &0.45 &  1.0 &1.3 &$\pm 2.8$   &$42\times 10^{-9}$    &$\pm 0.9$ & $ 0 < z_D < 2$\\
      3 &15 & 532 &0.12 & 11.8 &1.4 &$\pm 1.3$   &$ 1\times 10^{-9}$    &$\pm 9.2$ & $-4 < z_D < 6$\\
      \hline
     \multicolumn{10}{l}{$^a$ estimated from $w_D$}\\
     \multicolumn{10}{l}{$^b$ estimated from $\kappa$ according to (\ref{dl})} \\
     \multicolumn{10}{l}{$^c$ range of the possible distances given $F$}
    \end{tabular}
  \end{center}
\end{table}

A set of 25 home-made lock-in amplifiers -- based on TEXAS micro-controllers MSP430F2234 -- processed in parallel the signals of the photodiode matrix. Next, the demodulated signals were used to determine the phase of the same X-ray signal at the displacement start and end, which -- in order to maximize the noise rejections -- were set so as to coincide with integer optical orders, namely with zero crossings of the demodulated signals. Apart from an integer number of periods, which is not of interest, the 25 start-to-end phase shifts map the differences between the local measurements of the same displacement and, in turn, the wavelength profile across the beam. The lattice parameter value, that is, the period of the X-ray fringes, was used to calibrate the map. The accuracy of the lock-in amplifiers was verified by providing all the same photodiode signal; the scatter of the observed 25 wavelengths, near to $10^{-11}\lambda$, confirmed the expected capabilities.

\begin{figure}
\centering
\includegraphics[height=58mm]{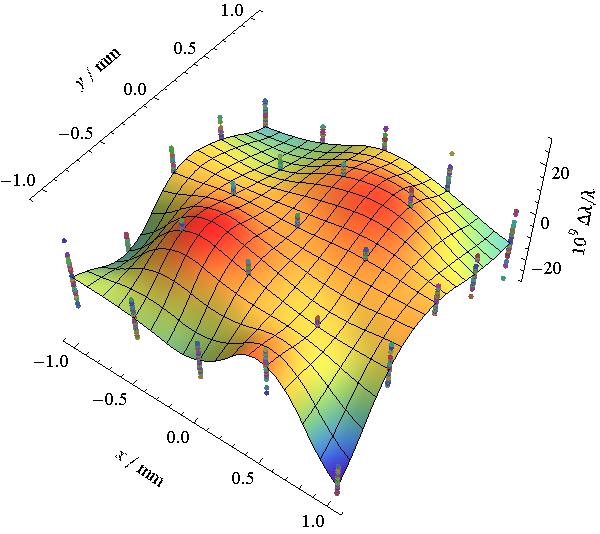}$\hspace{-5mm}^{1a}$\hspace{4mm}
\includegraphics[height=58mm]{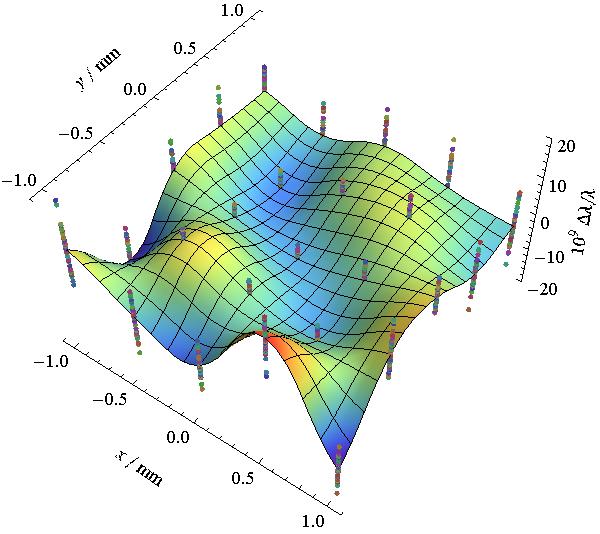}$\hspace{-5mm}^{1b}$\hspace{4mm}
\includegraphics[height=58mm]{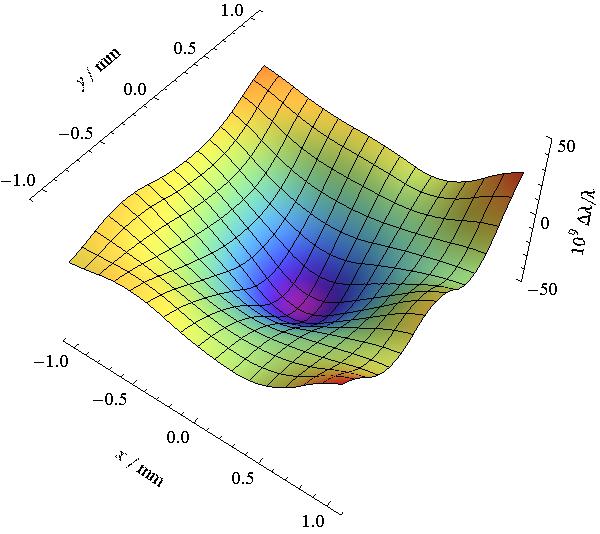}$\hspace{-5mm}^{2a}$\hspace{4mm}
\includegraphics[height=58mm]{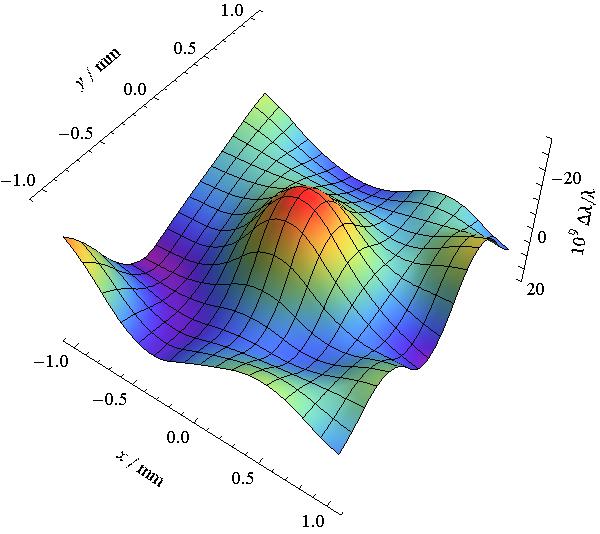}$\hspace{-5mm}^{2b}$\hspace{4mm}
\includegraphics[height=58mm]{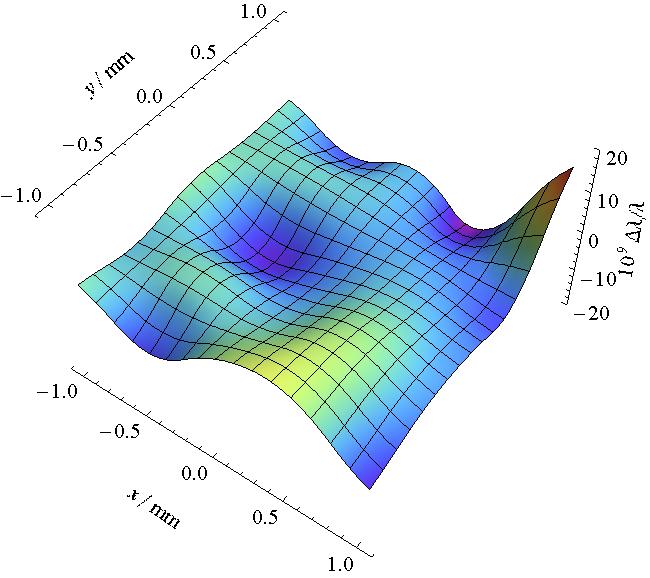}$\hspace{-5mm}^{3a}$\hspace{4mm}
\includegraphics[height=58mm]{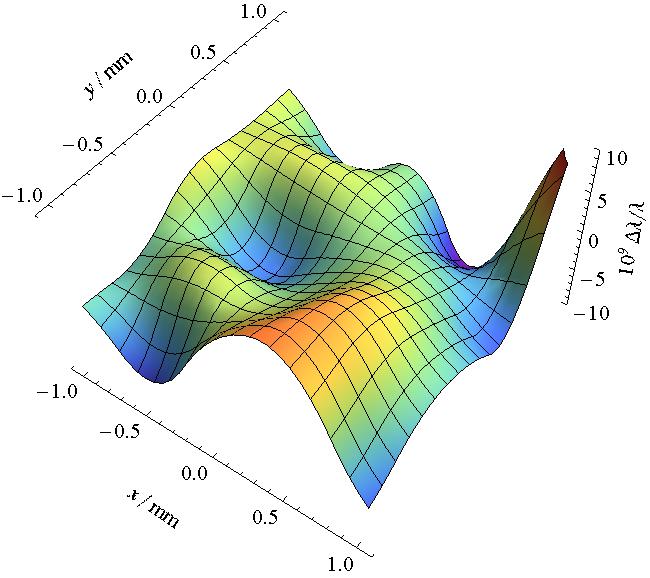}$\hspace{-5mm}^{3b}$\hspace{4mm}
\caption{Wavelength surveys across the beams whose parameters are given in table \ref{table:1}. Left: wavelength profiles. Right: residuals after the best-fit parabolas were removed. In $2b$, for the sake of clearness, residuals are shown upsidedown. The $(5\times 5)$ pixel$^2$ images were scaled down to take the 3$\times$ magnification into account. The first line shows the scatter of 50 subsequent profiles spaced by about 2 mm. The beam parameters are given in table \ref{table:1}.}\label{hama}
\end{figure}

Measurements were made with different laser beams, whose parameters are given in table \ref{table:1}. It compares also the detector distance from the beam waist estimated from the measured radius at the detection plane against the distance range estimated from the focal length of the fiber collimator and the detector distance from the collimator, about 1 m. The results of the wavelength surveys are shown in Fig.\ \ref{hama}. The right column shows the wavelength profiles, that is, the local distance between the interfering wavefronts. In the inset $2a$, the expected curvature is clearly visible. The curvatures of the best parabolas fitting the data are given in table \ref{table:1}. The left column of Fig.\ \ref{hama} shows the residuals after the best-fit parabolas were subtracted from the data and displays ripples that, if compared to the aimed measurement-accuracy, are quite large. Therefore, they deserve an explanation.

\subsection{Proposed explanation}
As it will be presently shown, the wavefront and intensity profiles show similar waves. Therefore, our conjecture is that the wavelength undulations originate from the wavefront evolution under paraxial propagation in the presence of irregular profiles. In the following, we outline a model of the interferometer operation and apply it to calculate how the wavefront waves drift away as the beam propagates and to investigate if their evolution explains an irregular wavefront separation.

\begin{figure}
\centering
\includegraphics[height=65mm]{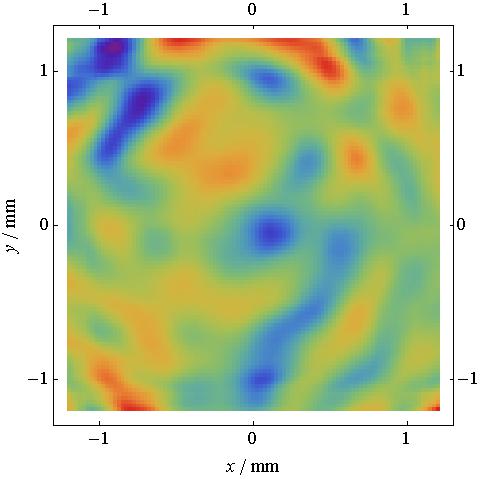}
\includegraphics[height=65mm]{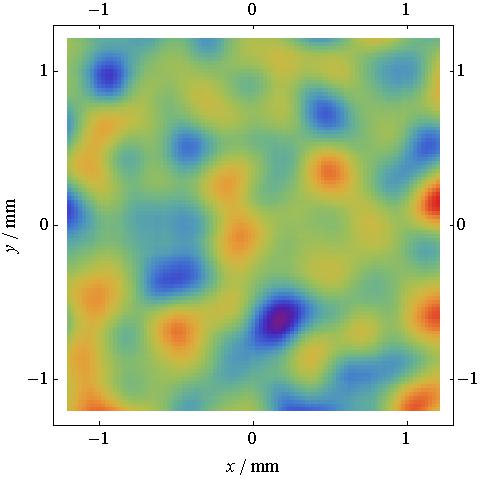}
\caption{Comparison of the observed and simulated phase noises. Left: observed phase-noise across the interfering beam \# 2 in table \ref{table:1}. Right: simulated phase noise. The color maps range from $-\lambda_0/30$ to $+\lambda_0/30$, the standard deviation is $\lambda_0/200$, the correlation length is 0.3 mm.}\label{pn}\vspace{2mm}
\includegraphics[height=65mm]{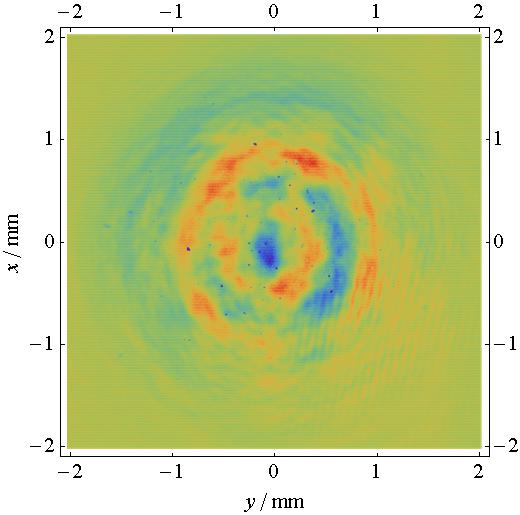}
\includegraphics[height=65mm]{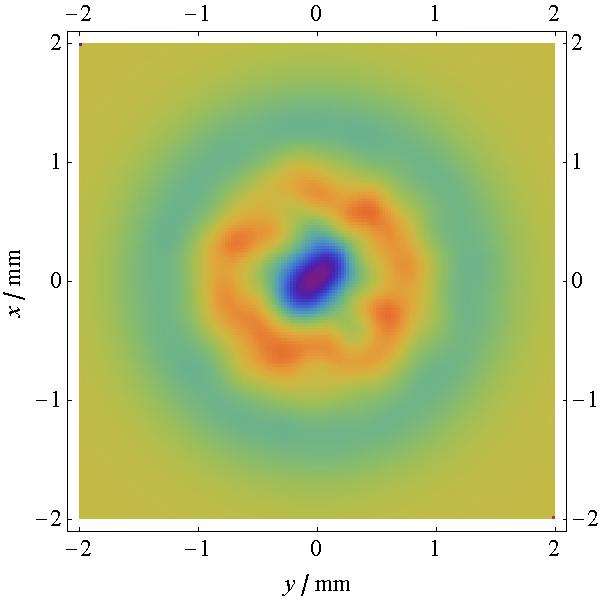}
\caption{Comparison of the observed and simulated intensity profiles. Left: residuals of the best Gaussian fit of the intensity profile of beam \# 2 in table \ref{table:1}. Right: simulated residuals. The color maps range from $-10\%$ (blue) to $10\%$ (red) of the maximum intensity.}\label{intensity}
\end{figure}

\subsubsection{Interferometer model}
To confirm that the observed ripples originate from the interferometer operation and to exclude unappreciated bugs in the experiment, let us consider the following model. In a two-beam interferometer, the laser beam is split in two parts (measure and reference) that recombine after propagating along the interferometer arms. Owing to the imperfect media and surfaces crossed, the split beams undergo different wavefront- and intensity-profile perturbations. We focus on those that originate before the path-length is varied, which diffract over a variable distance before being detected on the observation plane.

Let $E(x,y;0)$, where $x$ and $y$ are the detector coordinates, be the complex amplitude of the measure beam at the detection plane. With the use of the reciprocal-space representation, after the beam path is varied by $s$, the complex amplitude changes to
\begin{equation}\label{free}
 \tilde E(p,q;s) = U(p,q;s)\tilde u(p,q;0)\rme^{\rmi k_0 s} ,
\end{equation}
where
\begin{equation}\label{ft}
 \tilde u(p,q;0) = \frac{1}{\sqrt{2\pi}} \int_{-\infty}^{+\infty} u(x,y;0) \rme^{\rmi (px+qy)} \,\rmd x \rmd y .
\end{equation}
is the angular spectrum,
\begin{equation}\label{U}
 U(p,q;s) = \exp\left[- \frac{\rmi(p^2+q^2)s}{2k_0} \right]
\end{equation}
is the reciprocal-space representation of the free-space propagator of the paraxial approximation of the wave equation, and $k_0=2\pi/\lambda_0$ is the wave number.

After the path-length change, the direct-space representation of the complex amplitude is retrieved via the inverse Fourier transform of the propagated angular spectrum $u(p,q;s)$,
\begin{equation}\label{ift}
 u(x,y;s) = \frac{1}{\sqrt{2\pi}} \int_{-\infty}^{+\infty}  \tilde u(p,q;s) \rme^{-\rmi (px+qy)} \,\rmd p \rmd q ,
\end{equation}
and the phase shift of the interference pattern is
\begin{equation}\label{phase:2}
 \Delta \phi(x,y) = k(x,y) s = k_0 s + \arg\big[u(x,y;s)\big] - \arg\big[u(x,y;0)\big] .
\end{equation}
By introducing the effective wave number, $k(x,y)$, such that $\Delta \phi(x,y) = k(x,y)s$, the local wavelength is calculated as
\begin{equation}\label{lambda}
 \lambda(x,y) = \lambda_0 \bigg\{ 1 - \frac{\arg\big[u(x,y;s)\big] - \arg\big[u(x,y;0)\big]}{k_0s} \bigg\} .
\end{equation}

\subsubsection{Interferometer simulation}
To check if the proposed model explains the observation, we assume that the complex amplitude of the laser beam at the detection plane is
\begin{equation}\label{Esim}
 u(x,y;0) = \big[ 1+\alpha(x,y) \big] \exp \big[-r^{2a}/w_D^2 + \rmi k_0 r^2/(2R_D) + \rmi\phi(x,y) \big] ,
\end{equation}
where $r^2=x^2+y^2$, $w_D$ mm and $R_D$ are the 1/e spot-radius and the radius of curvature of the beam's wavefront at the detector, and $\alpha(x,y)$ and $\phi(x,y)$ are intensity and phase noises, respectively.

The beam parameters used in the simulation -- $\lambda_0=532$ nm, $\theta_0=0.45$ mrad, $z_R=2.8$ m, $w_D=1.3$ mm, and $R_D=3.0$ m -- have been chosen to correspond to the second entry of table \ref{table:1}. Since the beam radius at the detection plane is 1.3 mm, to avoid edges' effects, we used an expanded $(6.4\times 6.4)$ mm$^2$ window; the sampling period was 0.025 mm, corresponding to a $256\times 256$ grid. To enhance the phase excess, the path-length change was set to $s=3\times 10^4 \lambda_0$. The numerical accuracy was assessed by calculating the wavelength profile of a perfectly Gaussian beam and by comparing the result against the analytical expression (\ref{dl}). The residuals' standard deviation from the best-fit parabola (\ref{dl}) is $1.2\times 10^{-3}\Delta\lambda/\lambda_0$, the curvature error is $6\times 10^{-3} \kappa$.

\begin{figure}
\centering
\includegraphics[height=58mm]{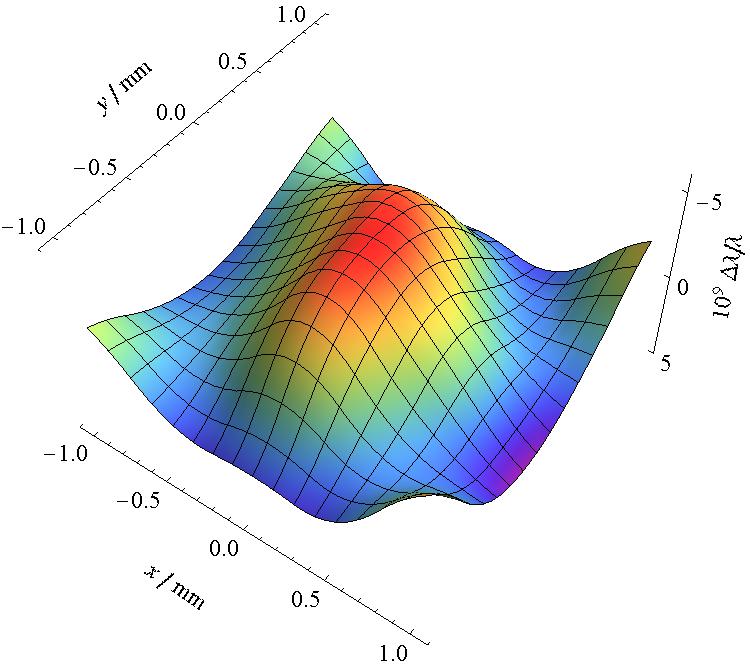}
\includegraphics[height=58mm]{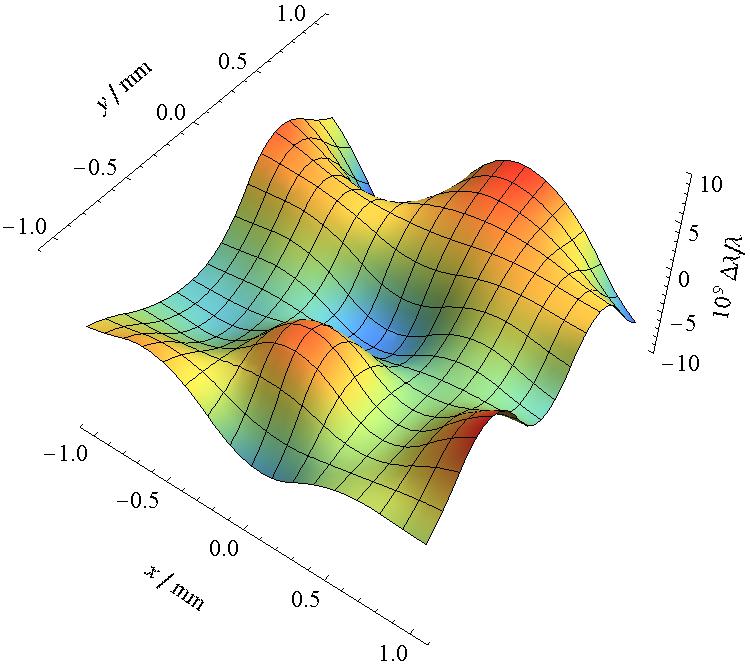}
\caption{Simulation results. Residuals after the best-fit parabolas were removed from the wavelength-profiles across the top-hat beam shown in Fig.\ \ref{intensity} (left, upsidedown) and a Gaussian beam where the noise shown in Fig.\ \ref{pn} was added to its phase (right).}\label{simulation}
\end{figure}

Because of the different media and surfaces crossed, the split beams undergo different perturbations, e.g., imprints of dust particles, surface roughness, and defects. Since the interference pattern at the output port delivers information on the $\phi(x,y)$ phase noise, we surveyed its inner $(2.5\times 2.5)$ mm$^2$ part from sequences of digitized intensity profiles grabbed while the path-length difference of the interfering beams is varied monotonically \cite{Balsamo:2003}. An example of the observed phase-noise is shown in Fig.\ \ref{pn} (left); the standard deviation and correlation length are $\lambda_0/200$ and 0.25 mm, respectively. To simulate the interferometer operation, we generated a random phase noise having zero mean and the same correlation length as the observed noise. A zoom of the simulated phase-noise over of the $(2.5\times 2.5)$ mm$^2$ inner area is shown in Fig.\ \ref{pn} (right).

As shown in Fig.\ \ref{intensity} (left), the beam intensity deviates from a Gaussian profile and displays ripples. To mimic the observed top-hat profile, the $a$ exponent in (\ref{Esim}) was set to 1.5 and $\alpha(x,y)$ was generated randomly with zero mean, 0.01 standard deviation, and 0.3 mm correlation length. A zoom of the simulated noise over of the $(4.2\times 4.2)$ mm$^2$ inner area is shown if Fig.\ \ref{intensity} (right).

The operation of the $5\times 5$ photodiode matrix was modeled by zooming the $(2.5\times 2.5)$ mm$^2$ detected part of the $(6.4\times 6.4)$ mm$^2$ simulation window and by averaging the wavelength calculated according (\ref{lambda}) over each of the 25 pixels, $(0.5\times 0.5)$ mm$^2$ wide. Since we substituted the average of (\ref{lambda}) for the average of the complex-amplitude (\ref{ift}), our simulation is not strictly correct. However, since the beam amplitude (\ref{ift}) varies slowly and its phase noise is small, this is an acceptable approximation. Figure \ref{simulation} shows the smoothed residuals after the best-fit parabolas were removed from the 25 wavelengths numerically obtained. The left inset refers to the top-hat beam shown in Fig.\ \ref{intensity} (right); no noise was added to the beam phase. The right one refers to a Gaussian beam having the noisy phase-profile shown in Fig.\ \ref{pn} (left). These results can be compared against the observation shown in Fig.\ \ref{hama}. Although the quantitative agreement is not perfect, the paraxial evolution explains the observed fluctuating separation of the wavefronts and the proper magnitude of the wave height. The scale difference between the profiles in Fig.\ \ref{hama}-$2b$ and the left inset of Fig.\ \ref{simulation}, is due to a "resonance" between the concentric noise-waves of the wavefront and intensity profiles, as shown in the left insets of Figs.\ \ref{pn} and \ref{intensity}.

\section{Conclusions}
A combined X-ray and optical interferometer was used to carry out wavelength surveys across a laser beam. Imprints were observed that correspond to $10^{-8}\lambda_0$ variations; they proved consistent with the paraxial evolution of wavefront perturbations and deviations from a Gaussian intensity-profile. When length measurements by laser interferometry aim at $10^{-9}$ relative accuracy, questions arise about their contribution to the measurement uncertainty. If the complex amplitudes of the interfering beams differ only by free-space propagation, the correction is known and depends only on the width of the beam angular-spectrum, no matter how irregular the wavefront and intensity profile may be. However, since the split beams undergo different perturbations, their difference cannot be traced back only to diffraction over different propagation distances. Further studies are under way to assess the effect of different imprints in the interferometer arms.

\section*{Acknowledgment}
This work was jointly funded by the European Metrology Research Programme (EMRP) participating countries within the European Association of National Metrology Institutes (EURAMET), the European Union, and the Italian ministry of education, university, and research (awarded project P6-2013, implementation of the new SI).

\end{document}